\documentclass[usenatbib]{mn2e}
\usepackage{graphicx,natbib,amssymb,aas_macros,lscape}
\def\degg{\hbox{$\null^\circ$\hskip-3pt.}}

\title[M31's Satellite Galaxies] {The satellite distribution of M31}
\author [McConnachie \& Irwin] {A.  W.  McConnachie$^{1,2}$ \& M.  J.
Irwin$^1$\\ $^1$Institute of Astronomy, Madingley Road, Cambridge, CB3
0HA, U.K.\\ $^2$Department of Physics and Astronomy, University of
Victoria, Victoria, B.C. V8P 5C2, Canada}

\begin{document}

\maketitle

\begin{abstract}
  The spatial distribution of the Galactic satellite system plays an
  important role in Galactic dynamics and cosmology, where its
  successful reproduction is a key test of simulations of galaxy
  halo formation. Here, we examine its representative nature by
  conducting an analysis of the 3-dimensional spatial distribution of
  the M31 subgroup of galaxies, the next closest system to our own. We
  begin by a discussion of distance estimates and incompleteness
  concerns, before revisiting the question of membership of the M31
  subgroup.  We constrain this by consideration of the spatial and
  kinematic properties of the putative satellites. Comparison of the
  distribution of M31 and Galactic satellites relative to the galactic
  disks suggests that the Galactic system is probably modestly incomplete at
  low latitudes by $\simeq 20\,\%$.  We find that the radial
  distribution of satellites around M31 is more extended than the
  Galactic subgroup; 50\,\% of the Galactic satellites are found
  within $\sim 100$\,kpc of the Galaxy, compared to $\sim 200$\,kpc
  for M31. We search for ``ghostly streams'' of satellites around M31,
  in the same way others have done for the Galaxy, and find several,
  including some which contain many of the dwarf spheroidal
  satellites. The lack of M31-centric kinematic data, however, means
  we are unable to probe whether these streams represent real physical
  associations. Finally, we find that the M31 satellites are
  asymmetrically distributed with respect to our line-of-sight to this
  object, so that the majority of its satellites are on its near side
  with respect to our line-of-sight. We quantify this result in terms
  of the offset between M31 and the centre of its satellite
  distribution, and find it to be significant at the $\sim 3\,\sigma$
  level. We discuss possible explanations for this finding, and
  suggest that many of the M31 satellites may have been accreted only
  relatively recently. Alternatively, this anisotropy may be related
  to a similar result recently reported for the 2dFGRS, which would
  imply that the halo of M31 is not yet virialised. Until such time as
  a satisfactory explanation for this finding is presented, however,
  our results warn against treating the M31 subgroup as complete,
  unbiased and relaxed.
\end{abstract}


\section{Introduction}

The Galactic satellite system consists of approximately a dozen
galaxies. This is an order of magnitude less than the number of haloes
predicted to exist as satellites from CDM simulations
(e.g. \citealt{kauffmann1993}), although this conflict is potentially
resolved by suppressing star formation in some subset of dark matter
haloes (e.g. \citealt{bullock2000}). Successful reproduction of the
Galactic satellite system, in terms of internal properties, frequency,
and spatial properties, is considered an important test of
cosmological galaxy formation models. In galactic dynamics, too, satellite
populations play an important role. For example, \cite{little1987},
\cite{kochanek1996}, \cite{wilkinson1999}, \cite{evans2000a} and
\cite{evans2000b} have used the Galactic and M31 satellite systems to
calculate the dynamical mass of the host galaxies.

\cite{willman2004} have questioned whether the current census of
Galactic satellites is complete, and suitable for use in the above
ways. They conclude that several satellites are probably missing at
low Galactic latitudes, and suggest that searches for Galactic
satellites could be suffering from incompleteness issues at large
Galactocentric distances. According to their study, these effects
could have resulted in a factor of up to 3 discrepancy between the
observed and actual number of Galactic satellites.

Around distant galaxies, a preference for satellites to lie along the
minor axis of the host galaxy has been reported, implying a preference
for polar orbits. This is the so-called Holmberg effect, first noted
by \cite{holmberg1969} and later revisited by \cite{zaritsky1997}.
However, recent results from SDSS question these findings, and
indicate a preference for orbits in the plane of the disk
(\citealt{brainerd2005}).  \cite{knebe2004} suggest these anisotropies
are the result of the accretion of satellites along filaments, leading
to a preference for satellites to be observed along the major axis of
the host halo and reflecting their initial infall onto the host
galaxy. However, given the ambiguous and contradictory nature of the
observational studies, the significance of this is not clear.

The Holmberg effect is not observed around nearby systems, but the
Galactic satellite distribution is anisotropic. The LMC and SMC are on
the opposite side of the sky to the Ursa Minor and Draco dwarf
spheroidals (dSphs), on an almost polar great circle on the sky.  On
the same great circle lies the Magellanic Stream
(\citealt{mathewson1974}), a large trail of HI gas which has been
stripped from the Magellanic Clouds.  This correlation in position led
\cite{lyndenbell1976} to hypothesise that the LMC, SMC, Ursa Minor and
Draco were once part of a ``Greater Magellanic Cloud'', which was torn
apart in the gravitational potential of the Galaxy and condensed to
form the latter two galaxies. The LMC and SMC are proposed to be the
surviving remnant of this body.

At around the same time, \cite{lyndenbell1982} noticed that the
Fornax, Leo I, II and Sculptor dwarf galaxies all appeared to align in
another polar great circle (the ``FLS'' stream), in a similar fashion to
the Magellanic Stream.  Later work by \cite{majewski1994} showed that
the then newly discovered Sextans and Phoenix dwarf galaxies both
aligned with this plane, along with most of the Galactic globular
clusters with the reddest horizontal branches. \cite{lyndenbell1995}
searched for all possible streams of satellites in the halo of the
Galaxy, including some of the distant globular clusters, and found
several candidates in addition to the Magellanic and FLS streams.

Kinematic evidence for these streams is still inconclusive. Only radial
velocities can be measured readily and although some proper motions
for these satellites exist, a common consensus has yet to emerge for
most of the dSphs. For example, \cite{schweitzer1996} conclude that
Ursa Minor appears consistent with it moving in the direction of the
Magellanic Clouds, although a more recent measurement by
\cite{piatek2005} concludes that its membership of the Magellanic
Stream is unlikely. Sculptor and Fornax have proper motion
measurements consistent with their membership of the FLS stream
(\citealt{schweitzer1996,dinescu2004}), although an earlier
measurement for Fornax had ruled out membership of this stream
(\citealt{piatek2002}).

\cite{kroupa2005} have suggested that the distribution of Galactic
satellites is inconsistent with the distribution of satellites
expected from CDM simulations. \cite{zentner2005}, however, point
out that a spherical distribution of satellites is not the correct
null hypothesis for dealing with CDM simulations, and the results of
\cite{kroupa2005} are marginally consistent with a prolate
distribution of satellites. \cite{libeskind2005} find that the subset
of subhaloes which have the most massive progenitors at early times,
and therefore arguably the ones most likely to be luminous, are
distributed in a similar way to the Galactic satellites, even though
the overall distribution of dark matter and dark satellites is very
different. In a similar way to \cite{knebe2004}, this result is
interpreted in terms of the accretion of satellites along filaments.

The satellite distribution of the Galaxy therefore plays an important
role in cosmology and galactic dynamics. Whether this distribution is
representative of satellite systems in general still needs to be
answered. As a result, the analysis of the spatial distribution of
similar systems is important to several fields. We are currently
involved in an extensive study of M31 and its environment
(\citealt{ibata2001a,ibata2004,ibata2005,ferguson2002,mcconnachie2003,mcconnachie2004a,mcconnachie2004b,mcconnachie2005a,mcconnachie2005b,chapman2005,irwin2005}).
This galaxy hosts the next closest satellite subsystem to our own and
in this contribution we conduct an analysis of its spatial
distribution and compare it to the Galactic satellites, to explore the
uniqueness of our satellite system. Some of the results presented here
have previously been discussed briefly in \cite{mcconnachie2004c} and
\cite{mcconnachie2005d}.  In Section~2 we discuss the initial set of
galaxies which we study, incompleteness concerns, and the distance
measurements that we use for each.  We also define a M31 spherical
coordinate system, in analogy to Galactic coordinates.  In Section~3,
we use the distance and radial velocity data for the Local group
galaxies to revisit the question of membership of the M31 subgroup.
In Section~4 we analyse the spatial distribution of the satellite
system and compare it with our own Milky Way.  In Section~5, we search
for ``ghostly'' streams around M31, in a similar way to
\cite{lyndenbell1995}. Section~6 discusses some of our findings, and
Section~7 summarises.

\section{The Data}

Most Galactic satellites are located within some 200\,kpc of the
Sun. Distances to these objects are reliably determined, and are
typically accurate to 5\%. It is therefore relatively easy to conduct
a reliable analysis of the distribution with respect to the Milky Way,
as the data is of high quality and only a small correction needs
to be applied to account for the offset of the Sun from the Galactic
centre.

M31 is located at a distance of $\simeq 780$\,kpc
(eg. \citealt{freedman1990,joshi2003,brown2004,mcconnachie2005a}). Some
14 to 18 other galaxies are believed to be associated with it at
roughly equivalent heliocentric distances (of order $600 -
1000$\,kpc). Accurate relative distances to each of these objects is
crucial for a study of their distribution around M31, as the
correction that must be applied to account for our offset from M31 is
two orders of magnitude larger than for the Galactic
centre. Additionally, it is important to have an understanding of any
possible incompleteness in the dataset (ie. currently undiscovered
group members). In this section, we first present the data that we
will be using, and define a coordinate system in which to analyse
them. We then discuss the accuracy of the data, and comment on
incompleteness effects.

\subsection{The Possible Members}

A full list of all galaxies which are initially considered as possible
members of the M31 subgroup are listed in Table~1, along with their
Galactic coordinates, adopted distance and uncertainty, and radial
velocity.  For completeness, we start by considering all Local Group
candidates in the same part of the sky as M31 that are not located
closer to the Milky Way than to M31. The majority of the distance
estimates are taken from \cite{mcconnachie2004b} (hereafter Paper~I)
and \cite{mcconnachie2005a} (hereafter Paper~II) and will be discussed
shortly. For those galaxies not analysed as part of our M31 survey
(objects listed below M32 in Table 1), we take their distances from
\cite{mateo1998a}.  M32 is assumed to lie at the same distance as
M31. The radial velocities for all objects other than Andromeda~IX are
those listed by the NASA/IPAC Extragalactic Database (NED). The radial
velocity of Andromeda~IX has been newly measured by \cite{chapman2005}
as part of the M31 radial velocity survey using the Deep Imaging
Multi-Object Spectrograph on Keck~II.

Table~2 lists the satellites of the Milky Way which will later be
compared to the M31 satellites. The distance information is taken from
\cite{mateo1998a}.  Leo~A is not considered a member of the Galactic
satellite system due to its relatively large separation from the Milky
Way. The putative dwarf galaxy in Canis Major (\citealt{martin2004a})
has not been included as its true nature is still the subject of some
debate. The ultra-faint Galactic companion in Ursa Major
(\citealt{willman2005}) is not included as it has only recently been
discovered and lacks a precise distance measurement.

\begin{table}
\begin{minipage}{0.475\textwidth}
\begin{tabular*}{1.\textwidth}{@{\extracolsep{\fill}}l r r c r}
\hline
\\ Galaxy & l & b & r (kpc)  & $v_\odot$ (km\,s$^{-1}$) \\
\hline\\
M31        &    121.2& -21.5&   785 $\pm$  25  & -301  \\
M33        &    133.6& -31.3&   809 $\pm$  24  & -180  \\
NGC\,205     &    120.7& -21.7&   824 $\pm$  27  & -244  \\
NGC\,147     &    119.8& -14.3&   675 $\pm$  27  & -193  \\
NGC\,185     &    120.8& -14.5&   616 $\pm$  26  & -202  \\
And\,I       &    121.7& -24.9&   745 $\pm$  24  & -380  \\
And\,II      &    128.9& -29.2&   652 $\pm$  18  & -188  \\
And\,III     &    119.3& -26.2&   749 $\pm$  24  & -355  \\
And\,V       &    126.2& -15.1&   774 $\pm$  28  & -403  \\
And\,VI      &    106.0& -36.3&   783 $\pm$  25  & -354  \\
And\,VII     &    109.5&  -9.9&   763 $\pm$  35  & -307  \\
And\,IX      &    123.2& -19.7&   765 $\pm$  24  & -216  \\
LGS3       &    126.8& -40.9&   769 $\pm$  23  & -286  \\
Pegasus    &     94.8& -43.5&   919 $\pm$  30  & -182  \\
WLM        &     75.9& -73.6&   932 $\pm$  33  & -116  \\
DDO210     &     34.0& -31.3&  1071 $\pm$  39  & -137  \\
M32        &    121.2& -22.0&   785 $\pm$  25  & -205  \\
IC10       &    119.0&  -3.3&   825 $\pm$  50  & -344  \\
IC1613     &    129.8& -60.6&   700 $\pm$  35  & -234  \\
EGB0427+63 &    144.7& -10.5&  1300 $\pm$ 700  &  -99  \\
\\
\hline\\
\end{tabular*}
\label{candidates}
\caption{All candidate members of the M31 sub-group initially
considered in this study, along with positional information and
heliocentric radial velocities.  M32 is assumed to be at the same
distance as M31. The three objects listed below M32 have had their
distances and uncertainties taken from Mateo (1998); all those objects
listed above M32 have had their distances and uncertainties taken from
Papers I \& II. The heliocentric radial velocities for all objects
other than Andromeda~IX are those listed by the NASA/IPAC Extragalactic
Database (NED).  Andromeda~IX was only been recently discovered by
Zucker et al.  (2004) and its radial velocity has newly been measured
by Chapman et al. (2005).}
\end{minipage}
\end{table}

\begin{table}
\begin{minipage}{0.475\textwidth}
\begin{tabular*}{1.\textwidth}{@{\extracolsep{\fill}}l r r c}
\hline
\\ Galaxy & l & b & r (kpc)  \\
\hline\\
LMC      &     280.5 &     -32.9 &  $~~49$ \\    
SMC      &     302.8 &     -44.3 &  $~~58$ \\  
Sagittarius &       6.0 &     -15.1 &  $~24~\pm~~2$ \\
Ursa\,Minor&     105.0 &     +44.8 &  $~66~\pm~~3$ \\   
Sculptor &     287.5 &     -83.2 &  $~79~\pm~~4$ \\   
Draco    &      86.4 &     +34.7 &  $~82~\pm~~6$ \\   
Sextans  &     243.5 &     +42.3 &  $~86~\pm~~4$ \\   
Carina   &     260.1 &     -22.2 & $101~\pm~~5$ \\  
Fornax   &     237.1 &     -65.7 & $138~\pm~~8$ \\  
LeoI     &     226.0 &     +49.1 & $250~\pm~30$ \\ 
LeoII    &     220.2 &     +67.2 & $205~\pm~12$ \\ 
Phoenix  &     272.2 &     -68.9 & $445~\pm~30$ \\ 
\\
\hline\\
\end{tabular*}
\caption{The satellites of the Milky Way and their Galactic
coordinates.  All distance information is taken from Mateo (1998).}
\end{minipage}
\end{table}

\subsection{M31-centric coordinates}

It is convenient to use a M31-centric coordinate system for this
investigation and we define a system that is directly analogous to
Galactic coordinates, but centered on M31. In Galactic coordinates, the
position of M31 is given by $\left(l_{M31}, b_{M31}, r_{M31}\right)$.
Another galaxy, $S$, has a position in Galactic coordinates of $(l, b,
r)$. The coordinates of $S$ in a Cartesian coordinate system, centered
on the Galaxy and aligned with $l_{M31}$, are

\begin{eqnarray}
x &=& r \cos\left(l - l_{M31}\right) \cos b \nonumber\\
y &=& r \sin\left(l - l_{M31}\right) \cos b \nonumber\\
z &=& r \sin b .
\end{eqnarray}

\noindent The coordinates of $S$ in the M31-centric spherical
coordinate system are $(l^\prime, b^\prime, r^\prime)$, and are given
by

\begin{eqnarray}
l^\prime&=&{\rm atan}\left(\frac{y^\prime}{x^\prime}\right) + 180^\circ \nonumber\\
b^\prime&=&{\rm asin}\left(\frac{z^\prime}{x^\prime}\right) \nonumber \\
r^\prime&=& \left(x^{\prime\,2} + y^{\prime\,2} + z^{\prime\,2}\right)^{\frac{1}{2}} 
\end{eqnarray}

\noindent where 

\begin{eqnarray}
\left( \begin{array}{c}
x^\prime + r_{M31}\sin i \\
y^\prime \\
z^\prime+ r_{M31} \cos i 
\end{array}\right)&
=&
{\mathbf R}_y \left(90 - i\right)
{\mathbf R}_x \left(90 - \theta\right)\nonumber\\
&&{\mathbf R}_y \left(-b_{M31}\right)
\left( \begin{array}{c}
x \\
y \\
z \\
\end{array}\right).
\end{eqnarray}

\noindent We take $\theta = 39\degg8$ to be the position angle of the
semi-major axis of M31 measured east from north in Galactic
coordinates, and $i = 12\degg5$ as the inclination of M31 to the
line-of-sight (\citealt{devaucouleurs1958}).  {\bf R} is the
appropriate Cartesian rotation matrix. The resulting coordinate system
is directly analogous to the Galactic coordinate system but is centered
on M31: $l^\prime$ is a longitude measured around the disk of M31,
where $l^\prime = 0$ is defined to be the longitude of the Milky Way,
and $b^\prime$ is a latitude, such that $b^\prime = 0$ corresponds to
the plane of the disk of M31.  $r^\prime$ is the distance of $S$ from
the centre of M31. The Galaxy therefore lies at $\left(l^\prime = 0,
b^\prime = -12\degg5, r^\prime = 785\,{\rm kpc}\right)$.  Henceforth,
the prime notation is dropped and the context will distinguish
Galactic and M31-centric longitude, latitude and distance.  The
uncertainty in the distance to each candidate satellite translates
into uncertainties shared by all three M31-centric coordinates, and
whose relative sizes depend upon the location of the individual galaxy
relative to M31.

\subsection{Distance measurements}

A study based on distances to multiple objects benefits from having
small random errors associated with each measurement. It is then
straightforward to account for the size of the error in the analysis
procedure and assess the confidence of any results. However, for these
methods to be reliable it is crucial that there are no systematic
differences between the measurements which remain unaccounted for by
the formal uncertainty.

Differential systematic uncertainties arise due to the heterogeneous
datasets that are employed, the standard candle utilised, the
observing/data reduction strategy and the algorithms applied to the
data. Distances in astronomy are historically unreliable, even for the
relatively nearby galaxies of the Local Group. It is important,
therefore, to use a homogeneous set of distances which have
differential systematic effects minimised. Any systematic
uncertainties that may be present will then, presumably, affect each
distance measurement in a similar way.

The I-band magnitude of the tip of the red giant branch (TRGB) is an
observationally and theoretically well determined standard candle for
relatively old, metal poor stellar populations. Using the Isaac Newton
Telescope Wide Field Camera (INT~WFC), we have obtained Johnson $V$
and Gunn $i$ photometry for a large number of Local Group galaxies,
the majority of which are members of the M31 subgroup. In Paper~I, we
developed an analysis procedure to minimise the observational
difficulties in measuring the location of the TRGB in a photometric
dataset. Paper~II then went on to apply this procedure to 17 Local
Group galaxies for which we have INT~WFC photometry and derived a
homogeneous set of distances, as well as metallicities, for these
objects. The identical data acquisition, reduction and analysis
procedure for each galaxy goes a large way towards minimising
differential systematic uncertainties, and the formal uncertainties on
the measurements are $\sim 5\,\%$, similar to the Galactic satellites.
The measurements are all consistent with previous estimates for each
of the targeted systems, and there is no evidence for any systematic
offsets or biases. The reader is referred to these papers for a more
thorough discussion of these measurements. For the purposes of this
study, there are a few systems that need to be considered for which we
have not derived distances in this manner and it is necessary to
supplement our data with that from other studies.

The distance to M31 is the key distance in this study. The distance
calculated for it in Paper~II is $785 \pm 25$\,kpc. This measurement
made use of stars in an annulus sampling the halo region of M31,
minimising the influence of the young stellar populations and the
substructure in this galaxy.  The annulus was far enough from the
centre of M31 so as not to be affected by crowding, significant
extinction or serious contamination from disk and bulge
components. Instead, it sampled a predominantly old, metal-poor
population and the TRGB was obvious as an abrupt change in star counts
in the luminosity function. Any line-of-sight effects through the halo
would be expected to smear out the TRGB, which was not observed (if
this effect was significant, then we would expect instead to have
measured the TRGB for stars in the halo of M31 which are closer to us,
and so the distance to M31 would be underestimated). The good
agreement with recent RR\,Lyrae ($794 \pm 37$\,kpc;
\citealt{brown2004}), Cepheid ($791 \pm 40$\,kpc; \citealt{joshi2003})
and independent TRGB distance measurements ($783 \pm 43$\,kpc;
\citealt{durrell2001}) suggests that our distance estimate is robust.

\subsection{(In)\,completeness of the M31 subgroup}
\label{completeness}

\begin{figure*}
\begin{center}
\includegraphics[width=6.5cm, angle=270]{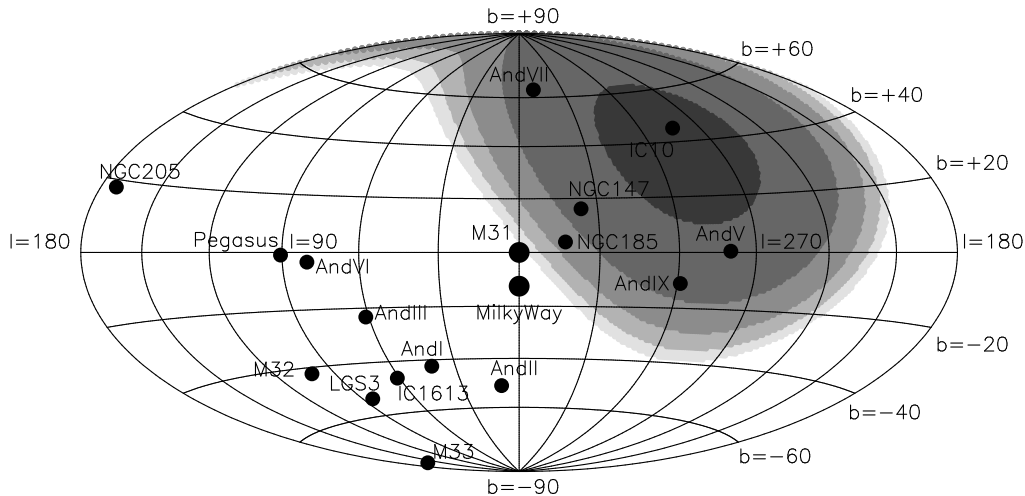}
\caption{The influence of the Galactic disk on the detection of M31
satellites, as seen in the M31-centric coordinate system. Contours
indicate the volume of space obscured by the Galactic disk. Objects
residing in the region of sky indicated by the central contour will
lie in the Galactic plane ($\left|b\right|<15^\circ$ in Galactic
coordinates) if they are $\ge 100\,$kpc from M31; objects lying inside
the next contour will lie in the Galactic plane if they are $\ge
150\,$kpc from M31, and so on out to $\ge 300\,$kpc from M31 (outer
contour). Shrinking of the contours by a factor of $2 - 3$ provides a
good representation of the region of sky covered by the innermost $b =
\pm 5^\circ$ of the Galactic disk, where we expect the extinction to be
most extreme.}
\label{extinction}
\end{center}
\end{figure*}

The pioneering survey by Sydney van den Bergh
(\citealt{vandenbergh1972b,vandenbergh1972a}) found the first dSph
companions to M31: Andromeda~I, II and III. This photographic (IIIaj)
survey was designed to detect objects fainter than the limits of the
Palomar Sky Survey, and was extended in 1973 -- 74 to encompass a
total area of $\sim 700$\,sq.\,degrees (\citealt{vandenbergh1974}). No
new dwarfs were discovered in this region. The limiting surface
brightness to which this survey is complete is unknown, and the sky
coverage becomes patchy beyond a projected distance from M31 of $\sim
100$\,kpc.  Subsequently, \cite{armandroff1998,armandroff1999} and
\cite{karachentsev1999} examined POSS~II IIIaj survey plates in the
vicinity of M31 and discovered three new companions to M31
(Andromeda~V, VI, VII).  During the same period,
\cite{whiting1997,whiting1999} conducted an all-sky search for new
Local Group galaxies. While this search found two new Local Group
galaxies, no new satellites of M31 were discovered.

The recent surveys for M31 satellites have been systematic in their
sky coverage surrounding M31 and the discovery of bodies such as
Andromeda\,V by \cite{armandroff1998} (with a central surface
brightness $\simeq 25.3$\,mags\,arcsec$^{-2}$;
\citealt{mcconnachie2005b}) show that they have been sensitive to low
surface brightness features. It seems probable that all the dwarf
galaxy companions to M31 with properties similar to those discovered
prior to 2004 have been discovered, due to the large ($\sim
1500$\,sq.\,degrees), contiguous nature of the surveys. Some
satellites, however, may lie directly behind the disk of M31. The area
of sky covered by the disk of M31 behind which satellites could not be
observed is $\sim 5 - 10$\,sq.\,degrees; only one or two satellites
are likely to hide from detection by this means.

The discovery of Andromeda~IX, with a central surface brightness in
$V$ of $26.8$\,mags\,sq.\,arcsec$^{-2}$ (\citealt{zucker2004a})
illustrates that the faint end of the M31 satellite luminosity
function is yet to be fully explored. It is currently impossible to
say if Andromeda~IX is unique, or whether it is the first of many
extremely low surface brightness M31 companions that await to be
discovered. In the hierarchical CDM framework, the luminous satellites
are expected to make up only a small subset of the total number of
dark matter haloes (\citealt{kauffmann1993}), with baryonic infall and
star formation being suppressed in the majority, perhaps by
reionisation or supernovae feedback
(eg. \citealt{bullock2000,dekel1986}). Cosmological simulations of
galaxy formation have difficulty modelling all these processes, and
star formation is usually dealt with by semi-empirical techniques. It
is important to constrain the relative importance of these effects by
obtaining an observational census of the number of these very faint
luminous satellites around the Galaxy and M31. Several searches using
SDSS are currently underway to do just this
(\citealt{zucker2004a,willman2005}). While these surveys are not yet
completed, the results so far indicate that there is not a significant
extra population of these objects.

To the north of M31 lies an area of sky obscured by the Galactic disk,
which could hinder the detection of M31
satellites. Figure~\ref{extinction} shows an equal-area Aitoff
projection of the satellites of M31 in the coordinates defined
previously. Due to the orientation of M31 with respect to
lines-of-sight through the Galactic disk, obscuration of M31
satellites is a function of position vector from M31. We consider the
Galactic plane to defined by $\left|b\right| < 15^\circ$ in Galactic
coordinates. Objects which reside in the region of sky indicated by
the central contour will lie in the Galactic plane if they are $\ge
100$\,kpc from M31; objects lying inside the next contour will lie in
the Galactic plane if they are $\ge 150\,$kpc from M31, and so on out
to $\ge 300$\,kpc from M31 (outer contour).

Several galaxies reside in the most badly affected quadrant and a few
of these, including one of the most intrinsically faint, lie in or
near the Galactic plane in areas of significant extinction, namely
And\,V, And\,VII, NGC\,185, NGC\,147 and IC10. For the first four of
these galaxies, the level of extinction is modest, at $0.4 -
0.7$\,mags in $V$. Only galaxies which were already intrinsically very
faint (comparable to And\,IX) might be hidden by this level of
extinction. On the other hand, IC10 is in a very heavily obscured
region of the sky and, according to the extinction maps of
\cite{schlegel1998}, has over 5\,mags of extinction in $V$. The exact
value is unreliable, given the patchy, irregular and large amount of
extinction very close to the plane of the galaxy. If other M31
satellites lie in these heavily-obscured regions, it is not clear that
they would have been easily observed; IC10 lends itself to discovery
as it is currently undergoing a starburst. If the contours in
Figure~\ref{extinction} are shrunk by a factor of 2 -- 3, they provide
a good representation of the area of sky covered by the innermost $b =
\pm\,5^\circ$ of the Galactic plane. This is the region of the
Galactic plane where we are most likely to get these extreme reddening
values. Considering the small area affected, it is unlikely that more
than a few satellites (at most) will be affected.

\section{Membership of the M31 subsystem}
\label{members}
\begin{figure}
\begin{center}
\includegraphics[width=7cm, angle=270]{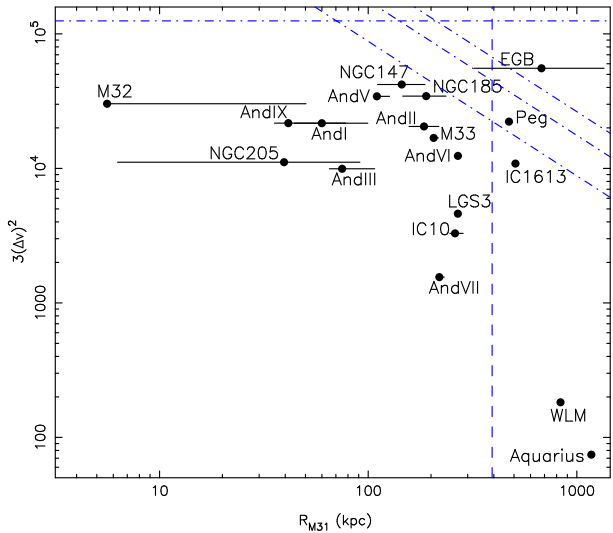}
\caption{The distance of each candidate M31 satellite from M31
compared with an estimate of the potential required for the satellite
to be virially bound to M31, as described in the text. Solid
horizontal lines represent the uncertainty in the separation of each
satellite from M31. The horizontal dot-dashed line represents the
escape velocity from an isothermal M31 halo of circular velocity
$\simeq 250$\,km\,s$^{-1}$.  The dotted lines correspond to the escape
velocity for a point mass of $1, 2$ and $3 \times 10^{12}$\,M$_\odot$.
The vertical dashed line is the distance of M31 from the barycentre of
the Local Group assuming the mass of M31 is of order the mass of the
Galaxy.  Galaxies lying in the area of the diagram that is
approximately delimited by these lines are most likely satellites of
M31.}
\label{lyndenbell}
\end{center}
\end{figure}

A useful definition for the membership of the M31 satellite system are
those objects whose kinetic energies are insufficient to escape the
gravitational potential of M31.  In addition, their separation must be
small enough such that it is more likely for them to be bound to M31
rather than the Local Group as a whole.  Choosing the correct
membership is complicated by uncertainties in the mass and extent of
the M31 halo, the separation of each candidate from M31 and the
velocity of each candidate relative to M31. However, the uncertainty
in the separation of the candidates from M31 is generally $\lesssim
\pm 40$\,kpc and is sufficient for this purpose. Likewise, the mass of
M31, although not known precisely, is of order 1 -- 2 $\times
10^{12}$\,M$_\odot$
(\citealt{evans2000a,evans2000b,ibata2004,fardal2005}) and is
sufficient to ascertain membership. The greatest practical uncertainty
in determining membership of the M31 subgroup comes from the velocity
estimate. Only the heliocentric radial velocity of M31 and its
companions can be measured directly and from this an estimate needs to
be made of their relative 3-dimensional velocity.

\cite{einasto1982} demonstrated that the tangential velocity of M31 is
small in comparison to the radial velocity under the assumption of no
net angular momentum in the Local Group.  Using this hypothesis, the
total Galactocentric velocity of M31, $\mathbf{v}_{M31}$, is well
approximated by the Galactocentric radial velocity component,
$v_{r,M31} = -123$ km\,s$^{-1}$.  We define the Galactocentric radial velocity
of $S$, located at a position $\mathbf{r}_s$ from the Galactic centre,
to be $v_{r,s}$. The component of the velocity of $S$ with respect to M31
in the direction $\hat{\mathbf{r}}_s$ is then

\begin{equation}
\Delta v = v_{r,s} -  v_{M31}\hat{\mathbf{r}}_s\mathbf{.}\hat{\mathbf{r}}_{M31}~,
\end{equation}

\noindent where $\hat{\mathbf{r}}_s$, $\hat{\mathbf{r}}_{M31}$ are
unit vectors. Independent velocity information for $S$ in the
remaining two orthogonal directions is unavailable. We therefore
assume equipartition of kinetic energy and multiply
$\left(\Delta\,v\right)^2$ by a factor of three.  The resulting
quantity is an estimate of twice the specific kinetic energy of the
galaxy, corresponding to an estimate of the gravitational potential
required for the putative satellite to be virially bound to M31. In
Figure~\ref{lyndenbell}, we plot this quantity against the separation
from M31 for each candidate satellite. The horizontal solid lines
indicate the uncertainty in the separation of each candidate from
M31. The horizontal dashed line corresponds to the escape velocity
from an isothermal halo with a circular velocity of $\simeq 250$
km\,s$^{-1}$.  The dotted lines correspond to the escape velocity for
a point mass of $1, 2$ and $3 \times 10^{12}$\,M$_\odot$. While these
mass profiles are unrealistic, they represent simple models which bracket the extremes of the global potential, without being too prescriptive. The
vertical line is the distance of M31 from the barycentre of the Local
Group, assuming that the masses of the Galaxy and M31 are roughly
equivalent. Objects lying in the area defined by these lines are
probably satellites of M31.

Figure~\ref{lyndenbell} shows that M32, NGC\,205, 147, 185,
Andromeda\,I, II, III, V, VI, VII and IX, LGS3, IC10 and M33 are
likely bound satellites of M31. The membership of Pegasus and IC1613
to the M31 subgroup is marginal as they are located $\sim 500$\,kpc
from M31. Nevertheless, we consider them to be members as their current
position relative to M31 and the Galaxy means that their orbits are
almost certainly dominated by the potential of M31. WLM and Aquarius
are located $\sim 1$\,Mpc from M31, and should not be considered as
satellites. The separation and velocity of EGB0427+63 means that it,
too, is unlikely to be a satellite.

\section{The 3D Distribution of the Andromeda Subgroup}

\begin{figure*}
\begin{center}
\includegraphics[width=9.cm, angle=270]{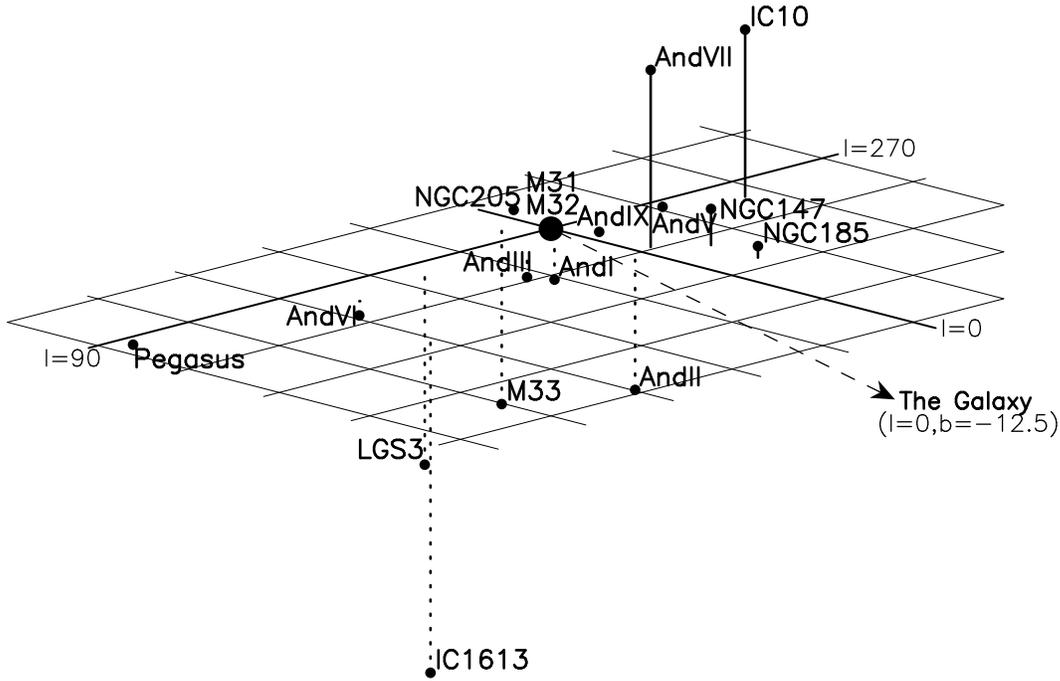}
\caption{A three-dimensional representation of the spatial
distribution of the M31 subgroup. The plane shown is aligned with the
disk of M31. Each grid cell is $100 \times 100$\,kpc. $l$ and $b$ are
the M31-centric spherical coordinates defined earlier. Solid lines are
used for satellites lying above the disk, while dotted lines are used
for satellites lying below the disk. The direction of the Galaxy is
marked.}
\label{3dpic}
\end{center}
\end{figure*}

\begin{figure*}
\begin{center}
\includegraphics[width=7.cm, angle=270]{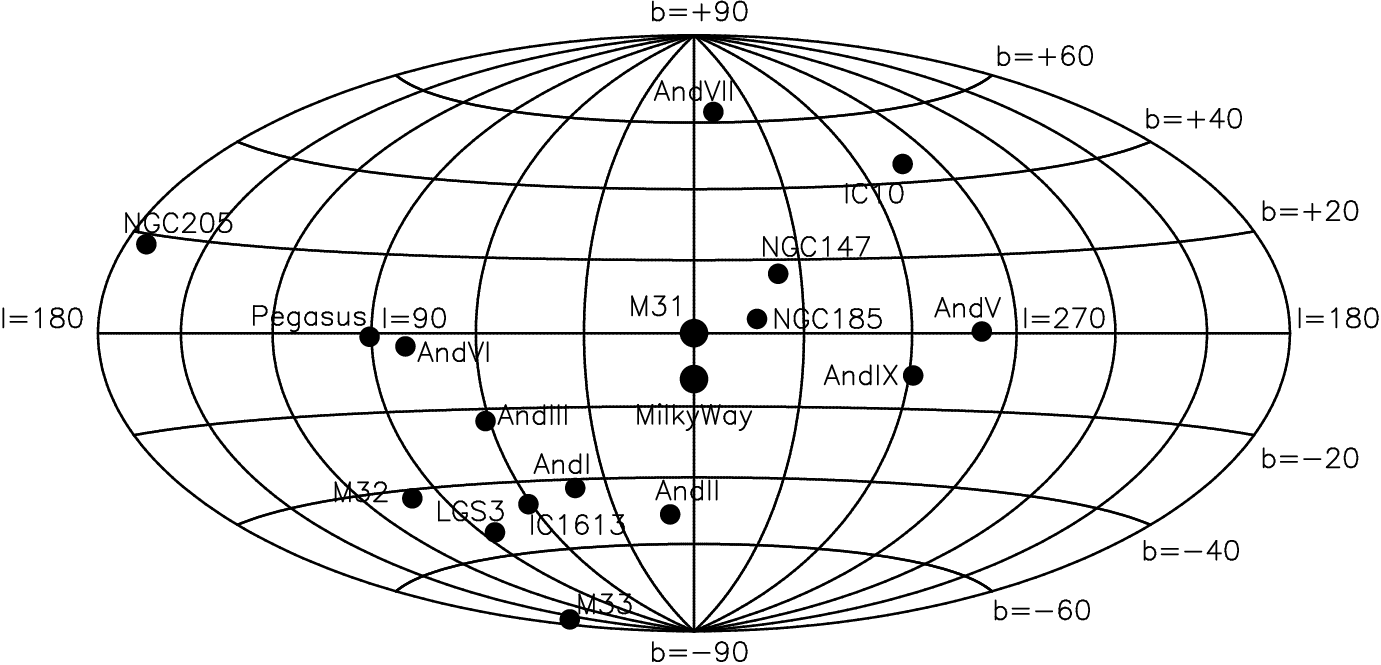}
\includegraphics[width=7.cm, angle=270]{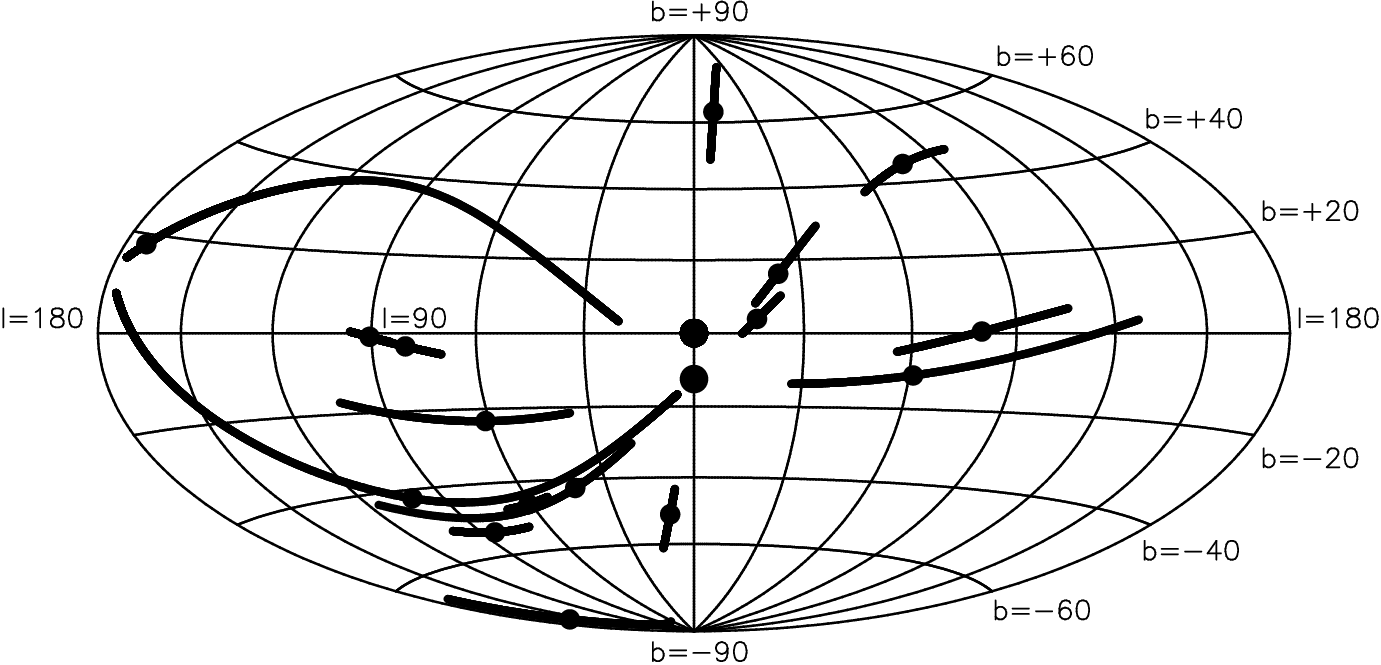}
\caption{Top panel: An Aitoff equal-area projection of the M31
satellite distribution, in the M31-centric spherical coordinates
defined earlier. The position of the Galaxy is shown for reference and
is defined to lie at $l = 0$. Bottom panel: Same as top, but now with
error bars showing the combined uncertainty in the position of each
satellite, due to the uncertainty in its distance and in the distance
to M31. The positions of NGC\,205, M32 and Andromeda\,IX are
relatively uncertain due to the close proximity of these objects to
M31. All of the other satellites occupy relatively well-defined
positions.}
\label{M31aitoff}
\end{center}
\end{figure*}

Figure~\ref{3dpic} shows the 3-dimensional distribution of the 16
members of the M31 subgroup.  The plane indicated is aligned with the
disk of M31 and each grid cell is $100 \times 100$\,kpc. Solid lines
are used for satellites that lie above the plane of the disk and
dotted lines are used for satellites that lie below the plane of the
disk. The direction of the Galaxy is marked. The top panel of
Figure~\ref{M31aitoff} shows the same distribution as an equal-area
Aitoff projection of the M31-centric $l$ and $b$ coordinates. The
position of the Galaxy is included as a reference.  The lower panel
shows error bars which represent the combined uncertainty in the
position of each satellite due to the uncertainty in its distance and
in the distance to M31. The positions of M32, NGC~205 and
Andromeda\,IX in this projection are sensitive to these uncertainties
due to their close proximity to M31, but the positions of the majority
of the satellites are relatively well constrained.

Several general points are clear from inspection of the top panel of
Figure~\ref{M31aitoff}. For example, half of the satellites lie within
$\left|b\right| \lesssim 20^{\circ}$ of the disk of M31. The satellite
distribution as a whole does not appear isotropic; the upper left
quadrant and lower right quadrant house only one satellite each, while
the remaining fourteen are confined to the other two quadrants. In
addition, the satellites appear to congregate within
$90^{\circ}$ of $l = 0$.

\subsection{Variation with latitude}

\begin{figure}
\begin{center}
\includegraphics[width=8.cm, angle=270]{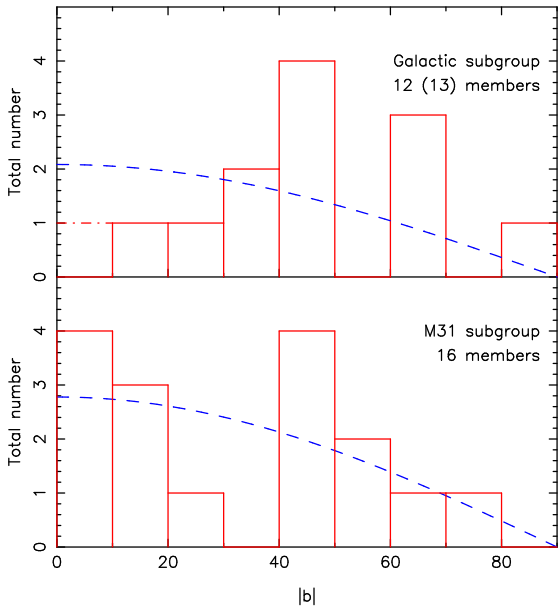}
\caption{The differential absolute latitude distribution of the
Galactic (upper panel) and M31 (lower panel) subgroups. The dashed
curves shows an isotropic distribution for the same total number of
satellites. Also shown as a dot-dashed line in the top panel is the
position of the putative Galactic satellite in Canis Major.  There is
an apparent deficit of Galactic satellites at low latitudes, which is
most readily explained as a selection effect due to obscuration by the
Galactic disk. The M31 and Galactic latitude distributions are not
statistically different however; a K-S test shows they are consistent
with being drawn from the same underlying distribution at the 38\,\%
level.}
\label{latitude}
\end{center}
\end{figure}

In the lower panel of Figure~\ref{latitude}, the differential absolute
(M31-centric) latitude distribution of the M31 satellites is
plotted. In the top panel, we show the Galactic satellite distribution
as a comparison (in Galactic coordinates). The putative dwarf galaxy
in Canis Major (\citealt{martin2004a}) has been highlighted in the
upper panel as a dot-dashed line. The dashed curves correspond to an
isotropic distribution of satellites. The correct null hypothesis for
a CDM distribution of satellites may not be isotropy, however, as the
satellites are expected to have a prolate distribution
(\citealt{zentner2005}). This makes no difference to the following
discussion.

With the inescapable caveat of small-number statistics, the M31
satellite distribution is qualitatively similar to the isotropic
distribution and suggests that isotropy is a reasonable first-order
assumption for the latitude distribution of satellites. Formally, a
Kolmogorov-Smirnov (K-S) test between the Galactic and M31
distributions shows that the two populations are not significantly 
statistically different. They are consistent with being drawn from the same
underlying distribution at the 38\,\% level. The inclusion of Canis
Major makes the distributions consistent at $> 60$\,\% probability.
However, the Galactic system has an apparent under-abundance of
satellites at low latitudes in comparison to the isotropic case and to
M31; only one (definite) satellite is observed around the Galaxy at
$\left|b\right| < 20^\circ$ (Sagittarius) compared to seven around
M31. If the Galactic satellites follow an isotropic distribution, then
we would expect $\sim 4$ satellites to be located at $\left|b\right| <
20^\circ$, instead of the 1(2) so far discovered. This implies that
obscuration by the Galactic disk at low Galactic latitudes may have
led to an incompleteness in the Galactic subgroup of $15 -
25$\,\%. While approximate, these values agree with the
study by \cite{willman2004}, who conclude that there may be a $\sim
33\,\%$ incompleteness in the Galactic satellites caused by
obscuration at low latitudes.

\subsection{Radial distribution}

\begin{figure*}
\begin{center}
\includegraphics[width=8.cm, angle=270]{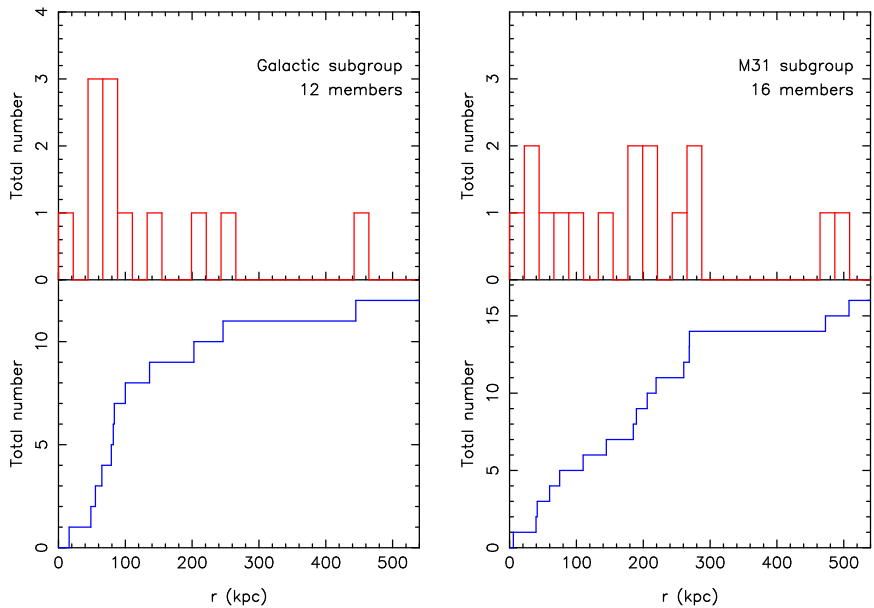}
\caption{The differential radial distribution of Galactic satellites
(top left) and M31 satellites (top right). The lower panels show the
respective cumulative distributions. The Galactic subgroup is more
centrally concentrated than M31; roughly half of its members are
located within 100\,kpc compared to $\sim 200$\,kpc for M31.  A K-S
test, where the populations are normalised to the same total number of
satellites, shows the distributions are not significantly statistically 
different; formally, there is a 24\,\% chance that they are drawn from the same
underlying distribution.}
\label{rad}
\end{center}
\end{figure*}

The top panels of Figure~\ref{rad} show the differential radial
distribution of Galactic satellites (left) and M31 satellites
(right). The lower panels show the respective cumulative
distributions. The abscissa have not been normalised by the virial
radii of the host haloes because these are uncertain and 
are believed to be similar for the two hosts (258\,kpc for the Galaxy,
280\,kpc for M31; \citealt{klypin2002}). The Galactic subgroup is
noticeably more centrally concentrated than the M31 subgroup; half of
the Galactic satellites are within $\sim 100$\,kpc while the
corresponding separation for the M31 system is $\sim
200$\,kpc. However, the distributions are not statistically different;
a K-S test on the distributions, normalised to the same total number
of satellites, shows there is a 24\,\% chance that they are drawn from
the same underlying distribution.

In a recent study, \cite{willman2004} compared a simulated CDM radial
distribution of satellites with those for the Galaxy and M31. These
distributions were compared by assuming that dwarfs inhabit the
sub-haloes with the highest circular velocity at zero redshift, and
assuming that the Galactic satellite system is incomplete beyond
100\,kpc. Based upon this comparison, they concluded that the Galactic
system is potentially incomplete by a factor of up to 3 (including the
effects of incompleteness at low Galactic latitude), and that there
may be some satellites beyond 100\,kpc which have eluded detection,
with absolute magnitudes and central surface brightnesses similar to
the detected population.

As the discovery of the ultra-faint stellar system in Ursa Major
illustrates (\citealt{willman2005}), it is almost certainly the case
that very faint satellites still await to be discovered around the
Galaxy. It is also likely that the Galactic disk hinders the detection
of low latitude Galactic satellites. It is not so clear, however, that
there is significant incompleteness in the outer ($> 100$\,kpc)
halo. \cite{irwin1994} conducted an automated search of Schmidt sky
survey plates for Galactic satellites, covering $\sim 2/3$ of the sky
and extending down to a Galactic latitude of $\left|b\right| =
20^\circ$. This survey was sensitive enough to detect systems such as
Draco and Ursa Minor out to $300$\,kpc. Only Sextans was found
(\citealt{irwin1990}). The survey could have found systems up to one
magnitude fainter than Sextans out to a distance of 200\,kpc. Despite
analysing more than 1000 sky survey plates, covering both hemispheres, no other
comparable satellites were found. In addition, given that the total
numbers of satellites around M31 and the Galaxy are roughly
equivalent, and that by normalising the populations by total number
the distributions are broadly similar, it seems unlikely that the
Galactic population is seriously incomplete in the manner suggested by
\cite{willman2004}.

\subsection{Distribution with respect to the Galaxy}

\begin{figure*}
\begin{center}
\includegraphics[width=7.cm, angle=270]{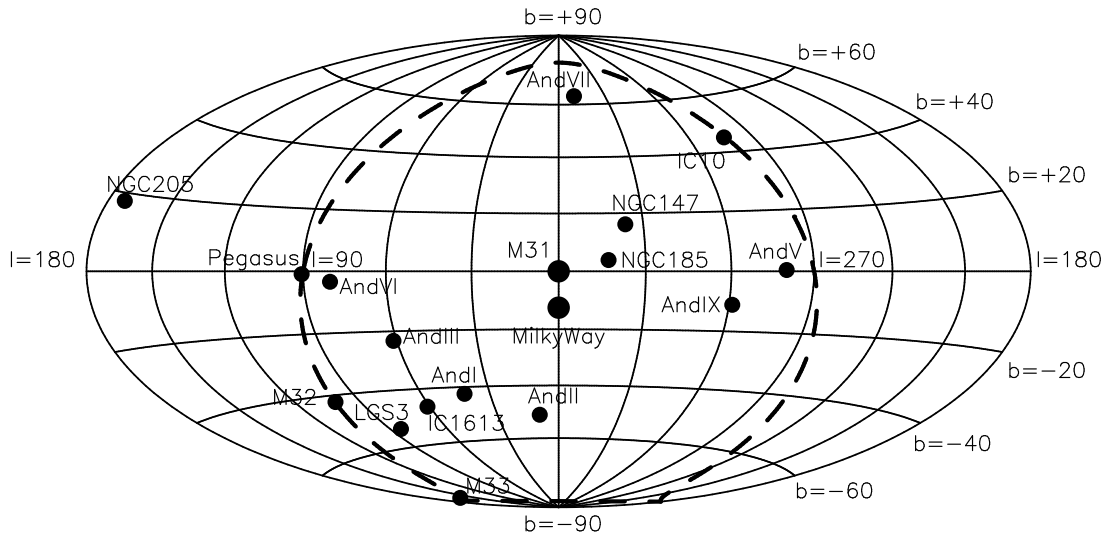}
\caption{Same as Figure~\ref{M31aitoff}, but with a dashed line
indicating the plane which separates the near side from the far side
of M31. The satellites of M31 show a correlation with this
plane, such that they preferentially lie on the near side of M31 with
respect to the Galaxy.}
\label{wonky}
\end{center}
\end{figure*}

Figure~\ref{wonky} is the Aitoff projection shown in
Figure~\ref{M31aitoff} with a dashed line representing the plane
separating the near side of M31 from the far side, with respect to the
Galaxy. Objects lying inside this line lie on the near side of M31,
while objects lying outside this line lie on the far side of
M31. Neglecting distance uncertainties, NGC205 is the only object that
clearly lies on the far side of M31; the other satellites are all
close to, or in front of, the plane. It is important to emphasise that
this plane is not a fit, and is defined only by the position of the
Galaxy with respect to M31. In fact, fourteen satellites lie in front
of the plane and two behind it. A suitable null hypothesis is that a
satellite is equally likely to lie in front or behind this plane. This
is easily tested with binomial statistics and is inconsistent with the
observed distribution at 99.8\,\%. This does not take into account the
distance uncertainty associated with each satellite. Therefore, we
have also adopted a Monte-Carlo sampling technique where we
continually generate new M31 satellite distributions by choosing the
distance of M31 and its satellites from Gaussian distributions
centered on the distance given for them in Table~1, with a standard
deviation equal to the corresponding distance uncertainty. Only in
0.5\,\% of the generated cases do we find a distribution with equal
numbers of satellites in both hemispheres, indicating that the
anisotropy is not an artifact of the distance uncertainties.
 
\begin{figure}
\begin{center}
\includegraphics[width=12.cm,angle=270]{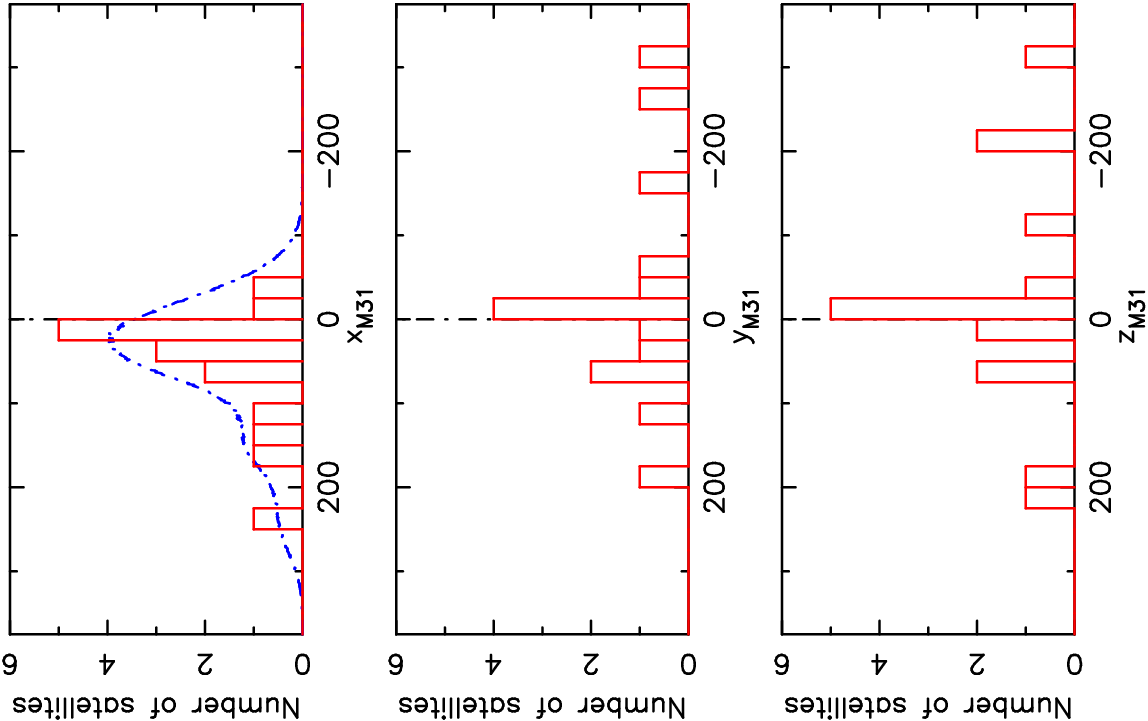}
\caption{Distribution of satellites along the line-of-sight to M31
(top panel) where $\left(x_{M31} = 0, y_{M31} = 0, z_{M31} = 0\right)$
is the position of M31 (dot-dashed vertical line) and $\left(x_{M31} =
785, y_{M31} = 0, z_{M31} = 0\right)$ is the position of the
Galaxy. The satellite distributions in the two directions orthogonal
to this (middle and bottom panels) are also shown. Distance
uncertainties mostly affect the distribution in $x_{M31}$. To
illustrate this, the blue dot-dashed curve shows the distribution of
satellites along the line-of-sight to M31, derived by continually
drawing the distance of M31 and its satellites from Gaussian
distributions centered on their derived distance with a standard
deviation equal to their distance uncertainty.}
\label{PDF}
\end{center}
\end{figure}

Figure~\ref{PDF} shows the positions of the satellites in a Cartesian
coordinate system, such that the abscissa is aligned with the
line-of-sight to M31. $\left(x_{M31} = 0, y_{M31} = 0, z_{M31} =
0\right)$ is the position of M31 and $\left(x_{M31} = 785, y_{M31} = 0,
z_{M31} = 0\right)$ is the position of the Galaxy. The distribution in
the top-panel is affected most by the distance uncertainties, and the
dot-dashed line represents the corresponding histogram for the
Monte-Carlo distributions generated previously.

For an isotropic distribution of satellites, the geometric centre of
the satellite distribution is expected to be coincident with the
position of the host. It is straightforward to calculate the mean (or median)
and random error in the mean ($\sigma = \sigma_d/\sqrt{N}$,
where $\sigma_d$ is the standard deviation) for the three
distributions shown in Figure~8. The random error in the
mean is the relevant statistic to use in this case, as individual
uncertainties in the position of each data-point statistically average out
when looking at the distribution of all data points (assuming these
uncertainties are of comparable magnitude and are random). As the
distributions deviate from Gaussians, we have calculated $\sigma_d$
using the more robust median-of-the-absolute-deviation-from-the-median
($MAD$) statistic (\citealt{hoaglin1983}). This is given by

\begin{equation}
MAD = median \left| x_i - median\left(x\right)\right|
\end{equation}

\noindent and is related to $\sigma_d$ by $\sigma_d = 1.48 \times MAD$
for a Gaussian. This statistic gives an estimate of the scatter in a
distribution even if it deviates significantly from a Gaussian and
contains several outliers. We have also calculated $\sigma_d$ using
the mean-of-the-absolute-deviation-from-the-mean statistic, finding
good agreement with the MAD statistic. 

\begin{table*}
\begin{center}
\begin{minipage}{0.9\textwidth}
\begin{tabular*}{1.\textwidth}{@{\extracolsep{\fill}}l r r r c r r r  }
&  \multicolumn{3}{c}{All satellites}  && \multicolumn{3}{c}{Excluding Peg + IC1613}\\
(kpc) & mean & median & $\sigma = \sigma_d/\sqrt{N}$ && mean & median & $\sigma = \sigma_d/\sqrt{N}$\\
$x_{M31}$: & 58 & 40 & 14 && 49 & 40 & 12\\  
$y_{M31}$: & -55 & -10 & 21 &&  -7 &  -5 & 19\\
$z_{M31}$: & -25 & -5 &  16 && -5 &  -4 & 17\\
\end{tabular*}
\caption{The mean, median, and random error in the mean ($\sigma$) for
the three distributions shown in Figure~\ref{PDF}.  These have been
calculated for the case where all the satellites are included and for
the case where the outlying Pegasus and IC1613 dwarfs are excluded.}
\label{stats}
\end{minipage}
\end{center}
\end{table*}

\begin{table*}
\begin{center}
\begin{minipage}{0.5\textwidth}
\begin{tabular*}{1.\textwidth}{@{\extracolsep{\fill}}r r r r}
(kpc)& mean & median & $\sigma = \sigma_d/\sqrt{N}$\\
$x_{M31}-785$: & 19 & 15 & 16\\  
$y_{M31}$: & 3 & -14 & 20\\
$z_{M31}$: & -43 & -26 &  21\\
\end{tabular*}
\caption{The same statistics as in Table~\ref{stats} but for the
Galactic satellite system. The coordinates are the same as defined
previously, but shifted to be centered on the Galaxy instead of
M31. No statistically significant offsets are observed between the
satellite distribution and the Galaxy.}
\label{MWstats}
\end{minipage}
\end{center}
\end{table*}

The mean, median and $\sigma$ for all the M31 satellites are listed in
Table~\ref{stats}. Also listed are the equivalent numbers when the
outlying satellites Pegasus and IC1613 are
excluded. Table~\ref{MWstats} lists the equivalent results for the
Galactic satellite system, where the coordinate system has been
shifted to be centered on the Galaxy. For this system, the random error
on the mean is of a similar magnitude, or greater, than the
mean/median of each coordinate. This suggests that there is no
statistically significant offset between the centre of the Galactic
satellite system and the Galaxy. Likewise, for the M31 subgroup, there
is no statistically significant offset between the centre of its
satellite system and M31 for $y_{M31}$ and $z_{M31}$ (taking into
account the effects of the two outliers). However, the mean/median of
$x_{M31}$ is inconsistent with zero at the $3\,\sigma$ level. This
suggests that the satellite distribution of M31 is significantly offset from 
M31 in the direction of the
Galaxy/barycentre of the Local Group at the $3\,\sigma$ level.

It is pertinent to calculate whether an equivalent asymmetry exists in
the velocities of the satellites.  Using the notation defined in
Section~\ref{members}, the mean velocity of the satellites relative to
M31 in the direction of the Galaxy is estimated as
$\left<\Delta\,v\right>/\left<\cos{\theta_{M31}}\right>$, where
$\cos{\theta_{M31}} =
\hat{\mathbf{r}}_s\mathbf{.}\hat{\mathbf{r}}_{M31}$. This is the
projection of the relative velocity of the satellite with respect to
M31 onto the line-of-sight to M31, and is $+19 \pm
21$\,km\,s$^{-1}$. A positive sign indicates that the satellites are
approaching more slowly than M31. The uncertainty is the random error
in the mean, and the velocity offset is not statistically significant.

\section{Ghostly Streams?}

\begin{figure*}
\begin{center}
\includegraphics[width=10cm, angle=270]{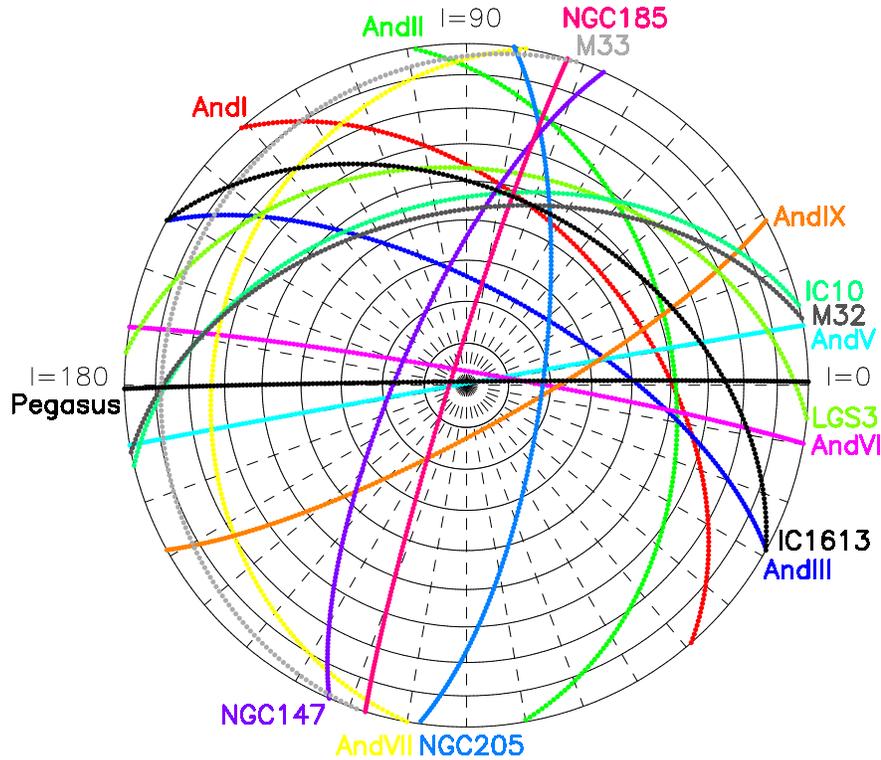}
\caption{A Lambert zenithal equal-area projection. $l$ increases
anticlockwise as shown. $b$ decreases radially from $b = 90^\circ$ at
the centre to $b = 0$ at the perimeter; the embedded circles represent
$10^\circ$ increments in $b$. Each curve represents the northern
`polar path' of the labelled M31 satellite ie. the track representing
all points tangential to its direction from M31. The point at which
two tracks cross defines the pole of the plane containing the two
objects and M31. Points at which multiple tracks cross define the
poles of possible streams containing the relevant satellites and M31.}
\label{lambert}
\end{center}
\end{figure*}

Using the 3-dimensional information available on the M31 satellite
distribution it is possible to search for ``ghostly streams'' of
satellites around M31 in a similar way to \cite{lyndenbell1995}. This
requires finding satellites that align on great circles, which may
reflect a stream-like origin and assumes that there has not been
significant precession of the orbits which would wash out any
alignments, as would be expected if the halo of M31 is
flattened. Here, we assume this is not the case. 

The probability of finding $n$ satellites from a sample of $N$ in one
particular plane is readily estimated from binomial statistics and is
small if $n$ is a large fraction of $N$. However, the probability of
finding $n$ satellites in any plane is much larger and reflects the
fact that all possible planes will have been tried in the search for
the plane around which the scatter is smallest. This point is relevant
to the claim by \cite{kroupa2005} that the planar distribution of
Galactic satellites is inconsistent with being drawn from a spherical
distribution. This is only true at a high significance level if the
plane in which the satellites align is considered to be predefined.

Figure~\ref{lambert} is a Lambert equal-area zenithal projection of
the polar paths of the satellites of M31. Each curve corresponds to
the track containing all points in the northern hemisphere
perpendicular to the vector connecting the satellite to M31. Where
each curve crosses another defines the pole of the plane containing
the two satellites and M31, and is equivalent to the direction
obtained on taking the cross-product of the position vector of the
satellites from M31. Points at which multiple paths cross define the
poles of planes which contain candidate streams of satellites.

Several possible streams are highlighted by Figure~\ref{lambert} but
attributing significance to them without kinematic information is premature. 
For example, the projected
position of M32 is sufficiently unreliable to neglect it in this
analysis, and NGC\,147 and 185 are probably a binary system
(\citealt{vandenbergh1998}), so should really be treated as a single
point.

However, there are numerous possible candidate streams indicated in
Figure~\ref{lambert}, of which some of the most well defined are:

\begin{itemize}
\item M33, Pegasus, IC10 (and possibly M32) $\left(l \simeq 180^\circ,
b \simeq 12^\circ\right)$;
\item Andromeda~III, VII and LGS3 $\left(l \simeq 140^\circ, b \simeq
23^\circ\right)$;
\item NGC\,185, 147 and 205 (and possibly Andromeda~II $\left(l \simeq
74^\circ, b \simeq 24^\circ\right)$;
\item Andromeda~I, IC10, 1613, NGC147 and 185 (and possibly M32)
$\left(l \simeq 80^\circ, b \simeq 42^\circ\right)$;
\item Andromeda~II, IC10 and LGS3 (and possibly M32) $\left(l \simeq
50^\circ, b \simeq 30^\circ\right)$;
\item Andromeda~III, V and IX $\left(l \simeq 10^\circ, b \simeq
56^\circ\right)$;
\item Andromeda~V, VI and Pegasus $\left(l \simeq 0, b \simeq
80^\circ\right)$. Andromeda~IX and NGC\,205 share a pole nearby at
$\left(l \simeq 0, b \simeq 72^\circ\right)$;
\item Andromeda~I, II, III, and VI $\left(l \simeq 352^\circ, b \simeq
38^\circ\right)$
\end{itemize}
 
Without doubt many, if not all, of these associations are chance
alignments; it is difficult to think of a plausible physical
association between M33, Pegasus, IC10 and M32, for example. However,
a few streams could be viable, such as those defined by the dwarf
ellipticals or some of the dwarf spheroidals. It is particularly
interesting to note the planes defined by the latter group;
Andromeda~I, II, III, V, VI and IX all share poles in the same general
vicinity of $\left(l \sim 0, b \sim 60^\circ\right)$ and the great
circle that they loosely form is readily identified in
Figure~\ref{M31aitoff}. These constitute six of the seven dSph
companions to M31 and have broadly similar star formation histories
and colours (Paper II). A physical association
between some or all of these objects may therefore be plausible, and
could represent analogous systems to the Magellanic or FLS streams
proposed around the Galaxy.

Further analysis of the planes listed above requires an investigation of
the kinematics of the candidate streams. The study of the Galactic
satellites by \cite{lyndenbell1995} investigated the specific
energy of the members of the proposed streams, under the assumption
that they were physically associated with each other and had the same
specific angular momentum, invariant with time. The derived specific
energies of the putative members are then expected to be equal if the
satellites are physically associated. The specific energy of each
satellite is given by

\begin{equation}
E = \frac{1}{2}v_G^2 + \frac{1}{2}h^2 r^{-2} - \phi~,
\label{speceng}
\end{equation}

\noindent where $v_G$ is the Galactocentric radial velocity of the
satellite, $h$ is the specific angular momentum and $\phi$ is the
gravitational potential of the Galaxy at the location of the
satellite. The first term represents the energy due to the radial
motion of the satellite; the second term is the rotational kinetic
energy of the satellite in its orbit, where $h$ has been assumed to be
a common constant for all stream members.

An equivalent kinematic analysis to that of \cite{lyndenbell1995} is
not directly possible for the M31 satellites. The heliocentric radial
velocities that are measured for these objects do not provide full
information on the M31-equivalent of $v_G$ and therefore the first
term of Equation~\ref{speceng} cannot be calculated. This will
remain the case until proper motions for the M31 satellites become
available. The additional uncertainties on the correct form and
magnitude of $\phi$, the distance uncertainties and the unknown
contribution of interlopers to the make-up of the candidate streams
(which may be as high as 100\,\%) also makes a numerical approach to
this problem unsatisfactory without prior information on the orbits of
at least some of the satellites. Without this information, any
conclusions on the physical association, or otherwise, of the members
of the M31 subgroup are likely to be highly vulnerable and uncertain.

\section{Interpretation}
\label{discuss}

Using the distances derived in Papers I and II, we have found that the
M31 satellites are distributed anisotropically along the line-of-sight
to this galaxy, and are skewed such that most satellites are on the
near-side of M31. We have quantified this distribution and measure a
$\ge 40$\,kpc offset between the centre of the M31 satellite system
and M31. This offset is in the direction of the Galaxy/Local Group
barycentre, and is significant at the $3\,\sigma$ level.

We have also examined the spatial distribution of the M31 subgroup
using earlier distance estimates to M31 and its satellites tabulated
in \cite{mateo1998a} (Andromeda~V, VI, VII and IX are recent
discoveries and are not included in this paper, so we kept their
distances the same). The larger distance errors in most of the other
measurements and the inhomogeneous nature of the distance estimates
which make up this compilation, unsurprisingly dilute, although do not
remove, the asymmetry observed in the distribution.

One of the primary motives of this paper was to make use of more accurate
and more consistent distance estimates to study the satellite distribution
of the M31 system.  These estimates are the key to
the subsequent analyses (see Section~2.3) and a comparison of them with respect
to previous distance estimates is given in Paper~II.  The
significant benefit obtained by using this dataset is that all of
the measurements are obtained on the same system and in this
respect represent the most homogeneous Local Group distance estimates
available. The distance estimate to M31 is clearly the most 
critical and bringing M31 closer by 25 -- 60\,kpc would 
remove the anisotropy. While this is a possibility, it requires the
TRGB measurement to M31 to be systematically offset relative to the
other TRGB measurements. The TRGB analysis, however, was designed to
minimise the probability of this occurring. The possible errors which
could affect M31 in particular were discussed in detail earlier. In
addition, this interpretation would require other TRGB
(\citealt{durrell2001}), RR\,Lyrae (\citealt{brown2004}) and Cepheid
(\citealt{joshi2003}) distances to M31 to be wrong in similar ways by
similar amounts. If M31 is located at $\sim 740$\,kpc, it is not obvious
why all the recent distance estimates to it, which have made
use of a range of independent standard candles, should produce very
similar wrong answers.  As such, we believe that this is the most
unlikely interpretation of the result.  It should not be neglected, but
it potentially has much wider reaching consequences for the use of these
distance indicators.

Selection effects were discussed in Section~\ref{completeness}. It is
hard to understand how this result could be explained by this
means. We are unaware of any effect which would prevent the detection
of M31 satellites located between $800 - 1000$\,kpc from the Galaxy,
assuming that they plausibly have typical properties of the satellite system. 
These satellites would have to be located directly behind the disk of M31 to
evade detection, and this is a very specific and small volume of space.
Only $\sim 5 - 10$\,sq.\,degrees of sky are masked by M31, compared to the 
$\sim 1500$\,sq.\,degrees over which satellites have been surveyed for and
discovered.

It is an interesting question to ask if such an offset could have a
dynamical origin. The main difficulty with a dynamical solution is
timescale: the orbital (dynamical) timescale of a satellite at a
distance of $\sim 100 - 200$\,kpc from a $\sim 10^{12}\,M_\odot$ host
is 1.6 -- 4.5\,Gyrs, meaning that satellites will have completed $>
3$\,orbits around M31 over the course of a Hubble time. Given this, it
is unlikely that any anisotropy present in the satellite distribution
could survive over a Hubble time, and suggests that it was either
introduced recently or maintained somehow. Further, any dynamical
effect is presumably going to affect both the luminous satellites and
the dark matter satellites, and so we would assume that the dark
matter satellites must also be asymmetrically distributed in this
manner.

A scenario in which a satellite spends a larger fraction of its
orbital period on the near side of M31 than the far side is easily
envisaged in a simple point mass case. Here, if the major axis of the
orbit of the satellite is aligned with the position vector of the
Galaxy, and M31 is at the far focus of the elliptical orbit, then the
satellite is more likely to be observed on the near side of M31, as
given by Kepler's laws.  In reality, the potential of M31 is not that
of a point mass and there is the added complication that the orbits of
the satellites need to be correlated. If the M31 subsystem consists of
dynamical groupings of satellites (Figure~\ref{lambert}), then the
probability of observing a heavily-skewed distribution might be
expected to increase, as there are fewer independent
satellite orbits. However, orbital precession in a non-spherical halo potential
would destroy any initial correlations present. This may either indicate that 
many of the dwarf satellites of M31 were accreted relatively recently (within 
the past dynamical time or so) or that the outer potential of M31 is 
approximately spherical.  There is a growing body of evidence
from simulations of galaxy formation that the observed dwarfs consist
of objects which were accreted relatively recently
(eg. \citealt{bullock2004,abadi2005}), else it is unlikely they could have
survived tidal disruption until the present. This may help explain the
skewed distribution.

Another possible dynamical explanation could be that the M31 satellite
galaxies are tracing the large-scale structure of the nearby
Universe. According to CDM structure formation, satellite galaxies are
accreted preferentially along filaments. It may be, therefore, that
the satellites were accreted along a filament that aligns with the
Galaxy/Local Group barycentre, or that the satellites are tracing a
dark filament that aligns with this direction. Alternatively, the
potential around M31 may be asymmetric due to contributions from
other Local Group components e.g. a large Local Group halo and/or a
contribution from the halo of the Galaxy, if it is significantly
extended. This latter scenario will have been particularly relevant at early
times when the separation of the M31 -- Galaxy binary system was much
less (the timing argument: \citealt{kahn1959,lyndenbell1981}) and if
the Galaxy is significantly more massive than M31. However, effects
such as precession could once again act to destroy any gross
asymmetries present.

\cite{vandenbosch2005} have recently shown from an analysis of the
Two-Degree Field Galaxy Redshift Survey (2dFGRS) that, in general, the
brightest galaxy in a group is offset in velocity from its satellites,
and has a specific kinetic energy that is $\sim 25\,$\% of its
satellites.  For a relaxed CDM halo, this corresponds to an offset of
the brightest galaxy from its satellite system of $\sim 3$\,\% of its
virial radius, or of order 10\,kpc for M31. The associated velocity
offset is a few tens of km\,s$^{-1}$. They suggest that the brightest
galaxy either oscillates around the minimum of the relaxed halo
potential (which is unlikely if the system has a cusp rather than a
core), or that it resides at the minimum of the halo potential which
is not yet relaxed.

It is unclear whether the scenarios developed by
\cite{vandenbosch2005} could apply to M31. These authors analysed
groups of galaxies defined by four or more members in the 2dFGRS picked out
using a group-finding algorithm developed by \cite{yang2005}. The
redshift range of these systems is $0.01 \le z \le 0.2$. They found
that for systems where the velocity dispersion was less than
$200\,$km\,s$^{-1}$ (corresponding to a total mass of $<
10^{13}\,M_\odot$), the central galaxy was consistent with being
located at rest at the centre of the group. However, for this mass
range, it was also consistent with being marginally displaced from the
centre. It may be, therefore, that the anisotropic satellite
distribution of M31 is tentative evidence that the halo of M31 is not
yet relaxed. To the best of our knowledge, there is currently no study which
gives the virialisation timescale of halos formed in a CDM hierarchy
as a function of mass.  The rather unusual distribution of satellites of M31
that we have discovered surely merits further theoretical studies.

\section{Summary}

We have re-derived membership of the M31 subgroup by examination of
the positions and kinematics of the putative members, and discussed
their distance estimates and completeness. Using an M31-centric
coordinate system, we have compared the distribution of satellites
above and below the disks of M31 and the Galaxy. This suggests that
the Galactic system is incomplete at low latitudes by $\sim
20\,\%$. Likewise, comparison of the radial distribution of satellites
around these two hosts show that the M31 system is less centrally
concentrated than the Galactic system.  We have also searched for
``ghostly streams'' of satellites around M31, in a similar way as
\cite{lyndenbell1995} did for the Galactic system, and find several,
including some which contain several of the dwarf spheroidal
companions. Unfortunately, the current lack of proper motion data for these
objects does not allow the possible physical association of these
satellites to be better constrained.

The analysis of the distribution of the M31 subgroup reveals that the
satellites are anisotropically distributed along our line of sight to
M31, and the satellite system as a whole is skewed relative to its host
in the direction of the Galaxy/Local Group barycentre. It is possible
that this result might be due to distance errors or selection effects,
but such explanations would require unlikely fine-tuning and the
former explanation in particular has potentially significant
consequences for the reliability and use of the main distance
indicators.  Some possible dynamical explanations are suggested but
present problems due to the timescales involved. It may be that many
of the M31 satellites were accreted into M31's halo only relatively
recently. Alternatively, it is possible that the asymmetric satellite
distribution that we observe is related to similar findings by
\cite{vandenbosch2005} at higher redshift, and that this may be
evidence to suggest that the halo of M31 is not yet
virialised. However, until an adequate explanation for this
distribution is presented, this finding warns against assuming that
the M31 subgroup represents a complete, relaxed and unbiased
population of galaxies.

\section*{Acknowledgements}
We would like to thank Rodrigo Ibata, Geraint Lewis, Annette Ferguson
and Nial Tanvir for help and discussions during the preparation of
this work. AWM would like to thank Mark Wilkinson, Justin Read, Neil
Wyn-Evans, Neil Trentham and Donald Lynden-Bell for enjoyable and
useful discussions of these results. Thanks also go to Beth Willman,
for her feedback on an earlier draft, and the anonymous referee, for a
very useful and thorough report.

\bibliographystyle{apj}
\bibliography{/pere/home/alan/Papers/references}

\end{document}